\documentstyle[preprint,aps]{revtex}

\begin{document}
\draft
\title{COMETARY KNOTS AND HIGH ENERGY PHOTONS AND NEUTRINOS 
FROM GAMMA RAY BURSTS} 
\author{Arnon Dar}
\address{Department of Physics and Space Research Institute\\
Technion - Israel Institute of Technology,\\ Haifa 32000, Israel.}

\maketitle

\begin{abstract} 

The Hubble Space Telescope has recently discovered thousands of gigantic
comet-like objects in a ring around the central star in the nearest
planetary nebula. We suggest that such circumstellar rings exist around
most stars. Collisions of the relativistic debris from gamma ray bursts in
dense stellar regions with such gigantic comet-like objects, which have
been stripped off from the circumstellar rings by gravitational
perturbations, produce detectable fluxes of high energy $\gamma$-rays and
neutrinos.

\end{abstract}




The isotropy of the positions of gamma ray bursts (GRBs) in the sky and
their brightness distribution as observed by the BATSE detector aboard the
Compton Gamma Ray Observatory (CGRO) strongly suggested [1] that GRBs are
at cosmological distances. The recent discovery [2,3] of an extended faint
optical source coincident with the optical transient of GRB 970228 and, in
particular, the detection of absorption and emission line systems [4] at
redshift z=0.835 in the spectrum of the optical counterpart [5] of GRB
970508, which may arise from a host galaxy, have provided further evidence
that GRBs are indeed at cosmological distances. Cosmological distances
imply [6] that GRBs have enormous fluences and peak luminosities. For
instance, in a Friedman Universe with $\Omega\approx 0.2$, $\Lambda=0$ and
$H_0\approx 70~km~Mpc~s^{-1}$ the luminosity distance of GRB 970508 is
$D=1.49\times 10^{28}cm$.  Its fluence [7] in the energy range $10-1000
keV~$ was $F_\gamma\approx 3\times 10^{-6} erg~cm^{-2}$ within a few
seconds, implying a total energy radiated in $\gamma$-rays of
approximately $(1+z)F_\gamma D^2d\Omega \approx 6\times
10^{50}d\Omega~erg$ where $d\Omega$ is the solid angle into which the
emitted radiation was beamed ($7\times 10^{51}~erg$ if the emission was
isotropic). Such intensities require that the gamma rays are highly
collimated in order to avoid self opaqueness due to $\gamma
\gamma\rightarrow e^+e^-$ pair production. A sufficient, and probably
necessary, condition for this to occur is that they are emitted far away
from the source by highly relativistic outflows with bulk Lorentz factors,
$\Gamma=1/\sqrt{1-\beta^2}\gg 1.$ Their non thermal spectrum also suggests
that they are emitted from highly relativistic flows, such as relativistic
fireballs [8] or relativistic jets [9]. The observed radiation may be
produced by self interactions within the flow [10] or by interactions with
external matter [11] or with external radiation [12].  If the highly
relativistic flow consists of normal hadronic matter, and if it collides
with gas targets along the line of sight before it decelerates
significantly, it will produce high energy $\gamma$ rays [13] and
neutrinos [14] through $pp\rightarrow \pi^0X$;  $\pi^0\rightarrow
2\gamma$, and $pp\rightarrow\pi^{\pm}X$; $\pi^{\pm}~\rightarrow
\mu^{\pm}\nu_\mu$; $\mu^{\pm} \rightarrow e^{\pm}\nu_\mu\nu_e $,
respectively, which are beamed towards the observer. Here we show that a
recent discovery with the Hubble Space Telescope implies that very large
number of such gas targets with a large column density may be present in
dense stellar regions, where GRBs are most likely to take place, and
produce detectable fluxes of high energy gamma rays and neutrinos from
GRBs. Although the Universe becomes opaque to $\gamma$-rays at TeV
energies due to $\gamma+\gamma_{_{IRB}}\rightarrow e^++e^-$ pair
production on the intergalactic IR background radiation [15], the
Universe is transparent to neutrinos (and to $\gamma$-rays below TeV),
which can be used to detect cosmological GRBs in TeV energies from
cosmological distances with $1~km^3$ underwater/ice neutrino telescopes
[16].

Recent observations with the Hubble Space Telescope of the Helix Nebula,
the nearest planetary nebula at an estimated distance of 150 pc, have
discovered [17] that the central star lights a circumstellar ring of about
3500 gigantic comet-like objects (``Cometary Knots'') with typical masses
about $M=1.5\times 10^{-5}M_\odot$ , i.e., comparable to our solar system
planets ($M_{Earth}=3\times 10^{-6}M_\odot$, $M_{Jup}=9.6\times
10^{-4}M_\odot$) and a total mass of about $M_{T}\sim 10^{-2}M_\odot$. The
gaseous heads of these Cometary Knots have a typical size of about
$r_c\sim 10^{15} cm$, which yields an average column density of $N_p\sim
M/\pi m_pr_c^2\sim 10^{22}~cm^{-2}$. They look like ionization bounded
neutral clouds, but it is not clear whether they contain a solid body or
uncollapsed gas.  They are observed at distances comparable to our own
Oort cloud of comets [18], but they seem to be distributed in a ring
rather than in a spherical cloud like the Oort cloud:  The Helix Nebula
has a structure of a ring lying close to the plane of the sky. The
Cometary Knots are not seen close to the central star, but become very
numerous as one approaches the inner ring and then become fewer out in the
main body of the ring (perhaps, because it becomes more difficult to
detect them there against the bright background).  This suggests a ring
rather than a spherical distribution, since no foreshortened objects close
to the direction of the central star are seen. It is possible that these
Cometary Knots have been formed together with the central star [17] since
star formation commonly involves formation of a thin planar disk of
material possessing too high an angular momentum to be drawn into the
nascent star and a much thicker outer ring of material extending out to
several hundreds AU [19]. Evidence for this material has been provided by
infrared photometry of young stars and also by direct imaging of this
material [20]. The observation of comets of very long period and 
unconfined to the ecliptic plane gave rise to the idea of the Oort cloud
of billions of comets in our Solar System at distances out to many tens of
thousands of AU [18]. It is argued that small gravitational perturbations
have circularized their parking orbits where occasionally another
perturbation puts them into an orbit which brings them into the inner
planetary system where they are finally viewable as comets. The mass
distribution of the observed comets has been determined over only a small
range since there are observational selection effects acting against
finding the small ones and there are very few large ones.  It is possible
that the Cometary Knots are the high mass end of the vastly more numerous
low mass comets [18] and are confined to the ecliptic plane because of
their relatively larger masses. It is also possible that most stars are
surrounded with such circumstellar rings of Cometary Knots. We propose
that these loosely bound Cometary Knots are stripped off from their mother
stars by gravitational perturbations in dense stellar regions (DSR), such
as star burst regions, super star clusters, galactic cores and core
collapsed globular clusters. Such DSR, where GRBs are likely to take place
[21], provide large column density gas targets for the highly relativistic
GRB debris. 
 
The cross section for inclusive production of high energy $\gamma$-rays
with a small transverse momentum, $cp_{T}=E_{T}<1~ GeV$ in pp collisions
is well represented by [22]
\begin{equation} 
{E\over \sigma_{in}} {d^3\sigma\over d^2p_{T}
dE_\gamma}\approx (1/2\pi p_{T}) e^{-E_{T}/E_0}~f(x), 
\end{equation} 
where $E$ is the incident proton energy, $\sigma_{in} \approx 35~mb$ is
the $pp$ total inelastic cross section at TeV energies, $E_0\approx 0.15~
GeV$ and $f(x)\sim (1-x)^3/\sqrt{x}$ is a function only of the Feynman
variable $x=E_\gamma/E$, and not of the separate values of the energies of
the incident proton and the produced $\gamma$-ray. The exponential
dependence on $E_{T}$ beams the $\gamma$-ray production into $\theta <
E_{T}/E$ along the incident proton direction. When
integrated over transverse momentum the inclusive cross section becomes
$\sigma_{in}^{-1}d\sigma/ dx\approx f(x).$ If the incident protons have a
power-law energy spectrum, $dI_p/dE\approx AE^{-\alpha}$, then, because of
Feynman scaling, the produced $\gamma$-rays have the same power-law
spectrum: 
\begin{equation} {dI_\gamma\over d E}
     \approx N_p \sigma_{in} \int_{E}^{\infty} {dI_p\over dE'}
      {d\sigma\over dE}dE' 
      \approx N_p\sigma_{in}IAE^{-\alpha},
\end{equation} 
where $N_p$ is the column density of the target and
$I=\int_0^1x^{\alpha-1}f(x)dx $.  Because of the exponential dependence of
the production cross section for $pp\rightarrow\gamma X$ on the transverse
energy of the produced $\gamma$-rays, most of the $\gamma$-rays that are
seen by the observer must arrive from within a cone of a solid angle
$\pi\theta^2 \approx \pi (E_0/E_\gamma)^2< d\Omega$ along the line of
sight.  The average column density within such a cone in a DSR, 
e.g., a galactic core [23] with a
typical radius, $R=R_{pc}~pc$, a total mass $M=M_6\times 10^6M_\odot$ and
a mass fraction $f=f_{-2}\times 10^{-2}$ in Cometary Knots [17], is
$N_{eff}\sim 3\times 10^{23}M_6f_{-2}R_{pc}~cm^{-2}.$ It produces a high
energy $\gamma$-ray flux of
\begin{equation} 
{dI_\gamma\over dE}
      \approx N_{eff}\sigma_{in}IAE^{-\alpha}.  
\end{equation} 
The duration of the high energy emission from such a cone is $T\approx
R/2c\Gamma^2+ R\theta^2/2c$. For $\Gamma\gg1/\theta\approx E_\gamma/E_0$,
\begin{equation} 
T\approx R\theta^2\approx 10^4R_{pc}E_{10}^{-2}~ s, 
\end{equation} 
where $E_\gamma=10E_{10}~GeV$.  Thus, we predict the ``afterglow'' of a
GRB in $\sim 10$ $GeV$ photons to extend over, typically, a couple of
hours and the typical duration of a pulse from a single cloud to last
$t\sim r_c^2/2cR\sim 5r_{15}^2/R_{pc}~s$. These predictions are consistent
with the CGRO/EGRET observations of the afterglow of GRBs in GeV photons
[24].
 
Hadronic production of photons in diffuse targets is also accompanied by
neutrino emission through $pp\rightarrow\pi^{\pm}X$;
$\pi^{\pm}~\rightarrow \mu^{\pm}\nu_\mu$; $\mu^{\pm} \rightarrow
e^{\pm}\nu_\mu\nu_e $.  If the incident protons have a power-law energy
spectrum, $dF_p/dE= AE^{-\alpha}$, and if the cloud is transparent both to
$\gamma$-rays and neutrinos, then because of Feynman scaling, the produced
high energy $\gamma$ rays and neutrinos have the same power law spectrum
and satisfy [25]: 
\begin{equation} dI_\nu/dE\approx
0.7 dI_\gamma/dE\approx 0.7 N_{eff}\sigma_{in}IAE^{-\alpha}. 
\end{equation}
Consequently, we predict that $\gamma$-ray emission from GRBs is
accompanied by emission of high energy neutrinos with similar fluxes,
light curves and energy spectra. The number of $\nu_\mu$ events from a GRB
in an underwater/ice high-energy $\nu_\mu$ telescope is $SN_AT_{GRF}\int
R_\mu(d\sigma_{\nu\mu}/dE_\mu)(dI_\nu/ dE)dE_\mu dE$, where $S$ is the
surface area of the telescope, $N_A$ is Avogadro's number,
$\sigma_{\nu\mu}$ is the inclusive cross section for $\nu_\mu p
\rightarrow \mu X$, and $R_\mu$ is the range (in $ gm~cm^{-2}$) of muons
with energy $E_\mu$ in water/ice. For a GRB with a fluence of $10^{52}
d\Omega~ erg$ in TeV $\gamma$-rays (similar to that observed in the
$10~keV-10~MeV$ band) and a power index $\alpha\sim 2$, as suggested by
the observed spectrum of GRBs in the multi $GeV$ energies [24], we predict
$\sim 1$ neutrino event in a $1~km^2$ telescope. Since the Universe is
transparent to neutrinos, they can be used to detect TeV GRBs from any
distance. In spite of the small number of events expected from individual
GRBs (except for a few relatively close), the fact that there are $\sim 3$
GRBs per day and the coincidence of neutrino events in time and direction
with GRBs can be used to enhance the cumulative neutrino signal from GRBs
over the atmospheric background by a very large factor. 

Hadronic production of TeV $\gamma$-rays is also accompanied by production
of TeV electrons and positrons mainly via $pp\rightarrow \pi^{\pm}X$;
$\pi^{\pm}\rightarrow \mu^{\pm}\nu_\mu$;  $\mu^{\pm}\rightarrow
e^{\pm}\nu_e\nu_\mu$.  Their production suddenly enriches the jet with
high energy electrons. Due to Feynman scaling, their differential spectrum
is proportional to the $\gamma$-ray spectrum \begin{equation}
dI_e/dE\approx 0.5 dI_\gamma/dE \end{equation} and they have the same
power-index $\alpha$ as that of the incident protons and the produced high
energy photons and neutrinos. Their cooling via synchrotron emission and
inverse Compton scattering from internal (trapped and produced) and
external (clouds, stars ) magnetic and radiation fields, respectively
[26], produce delayed emission of $\gamma$-rays, X-rays, optical photons
and radio waves with a differential power-law spectrum (assuming no
absorption in the cloud) \begin{equation} dI_\gamma/dE\sim
E^{-(\alpha+1)/2}, \end{equation} where $(\alpha+1)/2\approx 1.75\pm
0.25~.$ Hence, emission of TeV $\gamma$-rays is accompanied by delayed
emission (afterglows) in the $\gamma$-ray, X-ray, optical and radio bands. 
 
The peak emission of synchrotron radiation by electrons with a
Lorentz factor $\Gamma_e$ traversing a perpendicular magnetic field
$B_\perp(Gauss) $ in the flow rest frame which moves  with a 
bulk Doppler 
factor $\delta=(1-\beta
cos\theta)/\Gamma$ occurs at photon energy [27] $E_\gamma \sim 5\times
10^{-12} B_\perp\Gamma_e^2\delta ~keV $. The electrons lose $\sim
50\%$ of their initial energy by synchrotron radiation in
\begin{equation}
\tau_c\approx 5\times 10^8 \Gamma_e^{-1} B_\perp^{-2}\delta~s\approx 
1.2\times 10^3B_\perp^{-3/2}E_\gamma^{-1/2}\delta^{-1/2}~s.  
\end{equation} 
Consequently, the time-lag of synchrotron emission is inversely
proportional to the square root of their energy. It is small for
$\gamma$-rays and X-rays but considerable ($\sim hours$) for optical
photons and ($\sim$ days) for radio waves. 
The spectral evolution of the afterglow
is a convolution of the spectral evolution of the production of high
energy electrons and their cooling time. It is harder at the beginning
and softens towards the end of the glow.
Such a tendency seems to have been observed in GRBs.
The fading of the afterglow has a simple power-law dependence on 
time [28], $\sim t^{3(\alpha-1)/4}$ consistent with the observed 
fading of the X-ray and optical afterglows of GRB 
970228 [29,2,3] and GRB 970508 [30,4].

Gamma ray emission from GRB are usually interpreted as synchrotron
emission or inverse Compton scattering of highly relativistic electrons in
the ejecta. Efficiency considerations require that the ejecta be purely
leptonic. However, it is difficult to explain extended GeV $\gamma$-ray
emission from GRBs if the ejecta is purely leptonic. The main difficulty
is the fast cooling of electrons and positrons which are assumed to
produce the initial short keV-MeV GRB. Moreover, such a model does not
provide a satisfactory explanation for the very short time-scale
variability of GRBs (because of limited statistics the true short time
variability of GeV emission is not yet known). Instead, in this letter we
have assumed that the highly relativistic flow which produces the GRB
consists of normal hadronic matter.  We proposed that high energy
$\gamma$-rays and neutrinos are produced efficiently by the interaction of
the high energy particles accelerated by the highly relativistic flow with
gas targets of sufficiently large column density. Such gas targets may be
provided by gigantic comet-like objects, stripped off from the
circumstellar rings of the stars in the dense stellar region where GRBs
are more likely to take place. Thousands of such gigantic comet-like
objects have been discovered recently with HST in a ring around the
central star in the Helix nebula.  The simple properties of hadronic
production of high energy $\gamma$-rays, which are well known from lab
experiments, together with the properties of the gigantic Cometary Knots
can explain extended GeV emission from GRBs. The model also predicts a
simultaneous emission of TeV neutrinos and a delayed emission (afterglow)
in the $\gamma$-ray, X-ray, optical and radio bands with comparable
integrated energies.  Although further observations may provide more
supporting evidence for the hadronic nature of the ejecta, conclusive
evidence, probably, will require the detection of TeV neutrinos from GRBs,
perhaps by the 1 km$^3$ neutrino telescopes which have been proposed.

\noindent 
{\bf Acknowledgment}: The author would like to thank D. Fargion
and A. Laor for useful discussions.


\begin{references}

   
\item
See, e.g., C. A. Meegan et al., Nature, {\bf 355}, 143 (1992); 
G.J. Fishman and C. A. Meegan, Ann. Rev. Astr. Ap. {\bf 33}, 415 (1995). 

\item
J. van Paradijs et al., Nature, {\bf 386}, 686 (1997).

\item
K. C. Sahu et al., Nature,  {\bf 387}, 476 (1997).

\item 
M. R. Metzger et al., Nature {\bf 387}, 878 (1997).

\item
S. G. Djorgovski et al., Nature {\bf 387}, 876 (1997).

\item
For a review of cosmological models see C.D. Dermer 
and T. J. Weiler,  Ap. Sp. Sc.  {\bf 231}, 377 (1995).

\item
C. Kouveliotou et al., IAU Circ. 6600 (1997). 

\item 
B. Paczynski, Ap. J. {\bf 308}, L43 (1986); J. Goodman, Ap. J. 
{\bf 308}, L47 (1986).

\item

See, e.g., B. McBreen et al., Astr. Ap. Suppl. {\bf 97}, 81 (1993);
C. Dermer and R. Schlickeiser, AIP, {\bf 307} (1994);
J. Roland et al.,  Astr. Ap. {\bf 290}, 364 (1994); 
N. J. Shaviv and A. Dar, Ap. J. {\bf 447}, 863 (1995);
A. Dar, submitted to Ap. J. Lett. 1997 (astro-ph/9704187). 

\item  
B. Paczynski and G. Xu,  Ap. J. {\bf 427}, 708 (1994); 
M. J. Rees and P. Meszaros, Ap. J. {\bf 430}, L93 (1994). 

\item
M. J. Rees and P. Meszaros, Month. not. Roy. Astr. Soc. {\bf 258}, 41P  
(1992); P. Meszaros and  M. J. Rees, Ap. J. {\bf 405}, 278 (1993). 
J. I. Katz, Ap. J. {\bf 422}, 248 (1994).

\item
See, e.g.,  A. Shemi, Mon. Not. Roy. Astr. Soc. {\bf 269}, 1112 (1994); 
N. J. Shaviv and A. Dar, Ap. J. {\bf 447}, 863 (1995);
N. J. Shaviv and A. Dar, in {\it Neutrinos, Dark Matter and 
The Universe}, eds. T. Stolarcyk et al. (Editions Frontieres 1997) p. 338.

\item 
See, e.g., J. I. Katz, Ap. J. {\bf 432}, L27 (1994).

\item 
See, e.g., 
B. Paczynski and G. Xu,  Ap. J. {\bf 427}, 708 (1994); 
N. J. Shaviv and A. Dar, in {\it Neutrinos, Dark Matter and 
The Universe}, eds. T. Stolarcyk et al. (Editions Frontieres 1997) p. 338;
A. Dar in {\it Very High Energy Phenomena In The Universe} ed.
J. Tran Thanh Van (Editions Frontieres) in press; 
E. Waxman and J. N. Bahcall, Phys. Rev. Lett. {\bf 78}, 2292 (1997).

\item
F. W. Stecker, et al., Ap. J. {\bf 415}, L71 (1993).


\item 
See, e.g., F. Halzen, in {\it Very High Energy Phenomena In The
Universe} ed. J. Tran Thanh Van (Editions Frontieres) in press; L.
Moscoso, ibid; B. Price, ibid; I. Sokalski et al. ibid. Weak experimental
limits from much smaller scale ($\leq 10^{-3} km^2$ underground detectors
were published by Miller et al., Ap. J. {\bf 428}, 629 (1994);  Y. Fukuda
et al., Ap. J. {\bf 435}, 225 (1994) and R. Becker-Szendy et al., Ap. J.
{\bf 444}, 415 (1995). 

\item
C. R. O'Dell and K. D. Handron, Astr. J., {\bf 111}, 1630 (1996). 


\item
J. H. Oort, J. H., 1950, Bull. Astr. Inst. Neth. {\bf 11}, 91 (1950).


\item
Beckwith, S. V. W., 1995, in {\it Science with the VLT}, eds.
J. R. Walsh and T. J. Danziger (springer, Heidelberg), p. 53 


\item
O'Dell, C.R. and Wen, Z. 1994, Ap. J., {\bf 436}, 194 

\item  
See, e.g., N. J. Shaviv, Ph.D Thesis, Technion Report 
Ph-96-16,  (unpublished).

\item
G. Neuhoffer et al., Phys. Lett. {\bf 37B}, 438  (1971); 
H. Boggild and T. Ferbel, Ann. Rev. Nucl. Sci. {\bf 24}, 451 (1974);
T. Ferbel and W.R. Molzon, 1984, Rev. Mod. Phys., {\bf 56}, 181.


\item 
D. A. Allen, ``The Nuclei of Normal Galaxies'' (eds. R. Genzel \& A.I    .
Harris, 1994) p. 293.     

\item
K. Hurley et al., Nature {\bf 372}, 652 (1994);
B. L. Dingus, Ap. Sp. Sci., {\bf 231}, 187 (1995).  

\item 
See, e.g.,
A. Dar, and N. J. Shaviv, N. 1996, Astroparticle  Phys. {\bf 4}, 343
(1966). 

\item
L. Maraschi, et al., Ap. J. {\bf 397}, L5 (1992)
S. D. Bloom and A. P. Marsher, AIP {\bf 280}, 578  (1993);
C. D. Dermer and R. Sclickeiser, Ap. J. {\bf 415}, 418 (1993);
P. S. Coppi, et al., AIP {\bf 280}, 559 (1993);
C. D. Dermer and R. Sclickeiser, Ap. J. Suppl. {\bf 90}, 945 (1994);
M. Sikora et al.,  Ap. J. {\bf 421}, 153 (1994);
R. Blandford and A. Levinson, Ap. J. {\bf 441}, 79 (1995); 
S. Inoue and F. Takahara, Ap. J. {\bf 463}, 555 (1996).

\item
Rybicki, G.B. \& Lightman, A.P. 1979, {\it Radiative Processes in Ap.}
(N.Y. Wiley)

\item 
See, e.g., R. Wijers et al. Mont. Not. Roy. Astr. Soc., in press 1997
(astro-ph/9704153); A. Dar, submitted to Ap. J. Lett. 1997 (astro-ph/9704187). 
  
\item 
E. Costa et al., Nature, {\bf 387}, 783 (1997) and references therein. 

\item
E. Costa et al., IAU Circ. No. 6649 (1997); 
T. J. Galama et al., IAU Circ. No. 6655 (1997); 
L. Piro et al., IAU Circ. No. 6656 (1997); 
A. J. Castro-Triado et al., IAU Circ. No. 6657 (1997)
B. E. Schafer et al., IAU Circ. No. 6658 (1997);
P. J. Groot et al., IAU Circ. No. 6660 (1997); 
M. Garcia et al., IAU Circ. No. 6661 (1997);
M. Mignoli et al., IAU Circ. No. 6661 (1997); 
C. Chevalier and A. Ilovaisky, IAU Circ. No. (1997);
A. I. Kopylov et al., IAU Circ. Nos. 6663, 6671  (1997);
A. Fruchter et al., IAU Circ. No. 6674 (1997).

\end{references}
\end{document}